\documentclass[preprint,5p,fleqn,fleqnarray]{elsarticle}
\usepackage{amssymb,natbib}
\biboptions{sort&compress}                   
\journal{Physics Letters B}

\def\p{\mbox{\boldmath$\displaystyle\mathbf{p}$}}
\def\g{\mbox{\boldmath$\displaystyle\mathbf{g}$}}

\def\bv{\mbox{\boldmath$\displaystyle\mathbf{\varphi}$}}
\def\hp{\mbox{\boldmath$\displaystyle\mathbf{\widehat{\p}}$}}
\def\0{\mbox{\boldmath$\displaystyle\mathbf{0}$}}
\def\s{\mbox{\boldmath$\displaystyle\mathbf{\sigma}$}}
\def\J{\mbox{\boldmath$\displaystyle\mathbf{J}$}}
\def\K{\mbox{\boldmath$\displaystyle\mathbf{K}$}}
\def\x{\mbox{\boldmath$\displaystyle\mathbf{x}$}}
\def\gp{\gamma^\mu p_\mu}
\def\openone{\mbox{\boldmath$\displaystyle\mathbb{I}$}}
\begin{document}

\begin{frontmatter}

\title{Elko as self-interacting fermionic dark matter with axis of locality}

\author{D.~V.~Ahluwalia}
\ead{dharamvir.ahluwalia@canterbury.ac.nz}

\author{Cheng-Yang~Lee}
\ead{cyl45@student.canterbury.ac.nz}

\author{D.~Schritt} 
\ead{dsc35@student.canterbury.ac.nz}

\address{Department of Physics and Astronomy,  Rutherford Building\\
University of Canterbury, 
 Private Bag 4800, 
Christchurch 8020, New
Zealand}


\begin{abstract}
We here provide further details on the construction and properties  of mass dimension one quantum fields based on Elko expansion coefficients. We show that by a judicious choice of phases, the  locality structure can be dramatically improved. In the process we construct a  fermionic dark matter  candidate which carries not only an unsuppressed  quartic self interaction but also a preferred axis. Both of these aspects are tentatively supported by the data on dark matter.

\vspace{21pt}
\centerline{\textit{Journal Reference: Physics Letters B 687 (2010) 248-252}}
\end{abstract}


\end{frontmatter}

 \section{Introduction}  
 
If one wishes to treat Majorana spinors in their own right as four-component spinors, and not as Weyl spinors in disguise (or, as G-numbers), one must extend them in such a way that not only the $+1$ eigenvalue, under charge conjugation operator, but also the $-1$ eigenvalue is incorporated. This was the starting point of the Elko formalism, and the unexpected results, reported in references ~\cite{Ahluwalia:2004sz,Ahluwalia:2004ab}. It was recognised by the authors of these papers that the usual introduction of a Majorana mass term still leaves a problem with the free Lagrangian density,
and that to prevent the Dirac-type mass term from vanishing identically, one had to invoke a new dual. The mentioned problem is akin to the one mentioned by Aitchison and Hey~\cite[Appendix P]{Aitchison:2004cs}. However, the authors of the 
Elko formalism chose not to follow the Grassmannisation of the Majorana spinors. It is in this departure that several new results were obtained. Most unexpected of these was the mass dimensionality of the field.

The new dual appeared as an ad hoc construct in the mentioned works.
 Here we give a full justification for the introduction of the Elko dual.  Similarly, the locality structure investigated in the original papers failed to fully appreciate the necessity of certain phases in the expansion coefficients in a field operator.\footnote{The authors of the original Elko papers are not be too harshly criticised for these lapses as almost every textbook on quantum field suffers  from a similar neglect. Two notable exceptions are the recent classics by Weinberg and Srednicki~\cite{Weinberg:1995mt,Srednicki:2007qs}. The authors of the present communication acknowledge the insights gained from these monographs.}
Here we attend to that and learn of their dramatic effects on the locality structure.
 
 At present, the quartic self interaction, as well as a preferred axis in the dark sector,  are observationally favoured for dark matter candidates~\cite{Spergel:1999mh,Wandelt:2000ad,Ahn:2004xt,Balberg:2002ue,Land:2006bn,Samal:2008nv,Frommert:2009qw}. In this communication we provide an \emph{ab initio} evidence that both of these aspects are naturally present in the Elko dark matter.

To avoid confusion, we note that
spinors of the Elko formalism  have spawned an intense activity among a group of mathematical physicists and cosmologists~\cite{Boehmer:2007dh,Boehmer:2006qq,daRocha:2005ti,HoffdaSilva:2009is,daRocha:2008we,daRocha:2007pz,Boehmer:2009aw,Shankaranarayanan:2009sz,Boehmer:2008ah,Gredat:2008qf,Boehmer:2008rz,Boehmer:2007ut,daRocha:2007sd,Fabbri:2009ka,Wei:2010ad}. 
Similar to the work of Gillard and Martin~\cite{Gillard:2009zw}  the emphasis in this communication  is on the quantum fields, and not so much on the spinors.

 \section{Theory of self-interacting fermionic dark matter with axis of locality}
 
  In this section we outline the construction of two quantum  fields with Elko as expansion coefficients.  The full details shall appear in an archival paper elsewhere. 

\subsection{Notation}
  Let $\phi(\p)$ be a left-handed ($\ell$) Weyl spinor of
 spin one half.  Under a Lorentz boost, it transforms as $\phi(\p) =
 \kappa_\ell \phi(\0)$ where\footnote{The boost parameter $\bv =
 \varphi\hp$, in terms of energy $E$ and momentum $\p = p \hp$
 associated with a particle of mass $m$, is given by $\cosh(\bv) =
 E/m$ and $\sinh(\bv) = p/m$.  By $\s=(\sigma_1,\sigma_2,\sigma_3)$ we
 denote the Pauli matrices. The symbol $\openone$ represents an
 identity matrix, while $\mathbb{O}$ stands for a null matrix. Their
 dimensionality shall be apparent from the context. }
 \begin{equation} 
    \kappa_\ell = \exp\left(-\frac{\s}{2}\cdot\bv\right)  = 
    \varrho\left(\openone - \beta^{-1}{\s\cdot\p}\right), 
    \label{eq:boostL}     
    \end{equation}    
   with
 \begin{equation}
  \varrho:= \sqrt{ \frac{E+m}{2 m}},\; \mbox{and}\quad
  \beta := E+m 
\end{equation}

Here, the $\0$ is to be interpreted as $\p\vert_ {p\to 0}$, and not as
 $\p\vert_ {p=0}$. This restriction can be removed, if necessary (for example, by
 working in `polarisation basis' which then comes with its own subtleties).  We  choose $\phi(\p)$
 to belong to one of the two possible helicities: $\s\cdot\hp
 \,\phi_\pm(\p) = \pm\, \phi_\pm(\p)$.   Following
 Ref.~\cite{Ahluwalia:2004ab} note that, (a) under a Lorentz boost, $\vartheta
 \Theta \phi^\ast(\p)$ transforms as a right-handed ($r$) Weyl spinor,
 $\left[\vartheta \Theta \phi^\ast(\p)\right] = \kappa_r
\left[\vartheta \Theta \phi^\ast(\0)\right]$, with
     \begin{equation} 
     \kappa_r = \exp\left(+\frac{\s}{2}\cdot\bv\right)  = 
     \varrho\left(\openone + \beta^{-1}{\s\cdot\p}\right),
      \label{eq:boostR}
     \end{equation}
 where $\vartheta$ is an unspecified phase to be determined below, and $\Theta$ is
 Wigner's time reversal operator for spin one half,
 $\Theta\left[\s/2\right]\Theta^{-1} = - \left[\s/2\right]^\ast$; and
 (b) the helicity of $\vartheta\Theta \phi^\ast(\p)$ is {\em opposite} to that of
 $\phi(\p)$,
	\begin{equation} \s\cdot\hp \left[ \vartheta \Theta \phi_\pm^\ast(\p)
	\right] = \mp \, \left[ 
	\vartheta\Theta \phi_\pm^\ast(\p) \right].
	\end{equation}
 In terms of $\Theta( = -i \sigma_2)$, the charge conjugation operator in the $r\oplus\ell$ spinorial space
 reads
	\begin{equation}
	S(C)=\left(\begin{array}{cc}
	\mathbb{O} & i\Theta \\
	-i\Theta & \mathbb{O}\end{array}\right) K,\label{eq:cco}
	\end{equation}
 where $K$ is the complex conjugation operator.

 \subsection{Elko}
 
Elko abbreviates the German phrase \textbf{E}igenspinoren des
    \textbf{L}adungs\textbf{k}onjugations\textbf{o}perators.  
 The
 four-component {\em dual} helicity spinors
	\begin{equation}
	\chi(\p)= \left(\begin{array}{c}
 	\vartheta\Theta \phi^\ast(\p)\\
	\phi(\p)
	\end{array}\right),\label{eq:taup}
	\end{equation}
 become eigenspinors of the charge conjugation operator, i.e. Elko, with eigenvalues $\pm 1$ if the phase $\vartheta$ is set to  $\pm \, i$
	\begin{equation}
		S(C)\; \chi(\p)\Big\vert_{\vartheta=\pm i}
                = \pm \chi(\p)\Big\vert_{\vartheta=\pm i}. \label{eq:tau}
	\end{equation}
 We parameterise a unit vector along the momentum of a particle, $\hp$, as $(\sin\theta
 \cos\phi,\sin\theta\sin\phi,\cos\theta)$ and adopt phases so that at
 rest 
	\begin{eqnarray}
	\phi_+(\0)&=& \sqrt{m}\left(\begin{array}{c}
	\cos(\theta/2) e^{-i \phi/2}\\
	\sin(\theta/2) e^{i \phi/2}
	\end{array}\right),\label{eq:phiplus}\\
	\phi_-(\0) &=& \sqrt{m}\left(\begin{array}{c}
	-\sin(\theta/2) e^{-i \phi/2}\\
	\cos(\theta/2) e^{i \phi/2}
	\end{array}\right). \label{eq:phiminus}
	\end{eqnarray}
 Equations~(\ref{eq:phiplus}-\ref{eq:phiminus}), when coupled
 with Eq.~(\ref{eq:taup}),
 allow us to explicitly introduce the self-conjugate spinors ($\vartheta=+i$) and 
 anti self-conjugate spinors ($\vartheta=-i$) at rest
	\begin{eqnarray}
	\xi_{\{-,+\}}(\0) &:=& +\; \chi(\0)\big\vert_{\phi({\bf{0}})
	\to\phi_+({\bf{0}}),\;\vartheta=+i,} \\
	\xi_{\{+,-\}}(\0) &:=& +\; \chi(\0)\big\vert_{\phi({\bf{0}})
	\to\phi_-({\bf{0}}),\;\vartheta=+i,} \\
	\zeta_{\{-,+\}}(\0) &:=& + \;\chi(\0)\big\vert_{\phi({\bf{0}})
	\to\phi_-({\bf{0}}),\;\vartheta=-i,} \\
	\zeta_{\{+,-\}}(\0) &:=& -\; \chi(\0)\big\vert_{\phi({\bf{0}})
	\to\phi_+({\bf{0}}),\;\vartheta=-i.} 
	\end{eqnarray}
 The $\xi(\p)$ and $\zeta(\p)$ for an arbitrary momentum are now
 readily obtained~\footnote{The boost operator commutes with the
 charge conjugation operator and for that reason $S(C)\; \chi(\0) =
 \pm \chi(\0)$ implies $S(C)\; \chi(\p) = \pm \chi(\p)$.}
	\begin{equation}
	\xi(\p) = \kappa\, \xi(\0),\quad\zeta(\p) = \kappa\, \zeta(\0),
	\end{equation}
where
	$\kappa:=
	\kappa_r \oplus
	\kappa_\ell
	$. The choice of phases and the dual-helicity designations are different
 from those adopted in
 references~\cite{Ahluwalia:2004sz,Ahluwalia:2004ab}. These changes were inspired by the considerations presented in Sec. 38 of
 reference~\cite{Srednicki:2007qs}, and by those given in Sec. 5.5 of reference~\cite{Weinberg:1995mt}).  These differences are crucial to the results here presented. 

\subsection{Elko dual}

If one now invokes the Dirac dual for the $\xi$ and $\zeta$ spinors one immediately encounters a problem
 in constructing a Lagrangian
 description~\cite[Appendix P.1]{Aitchison:2004cs}.
 This was one of the reasons that a new dual was introduced in the original papers on Elko. That dual translates to the following definition
		\begin{equation}
		{{\stackrel{\neg}{e}}_{\{\mp,\pm\}}(\p)} :=  \mp i
		\left[e_{\{\pm,\mp\}}(\p)\right]^\dagger \gamma^0.
		\label{eq:dualchi} \\
		\end{equation}
Its essential uniqueness can be established by looking for 
a `metric' $\eta$ such that the product $
[e_\imath (\p) ]^\dagger \eta\, e_\jmath(\p)  $ \textemdash~with $e_\imath(p)$ as any one of the four Elko  \textemdash~remains invariant under 
an arbitrary Lorentz transformation. This requirement can be readily shown to translate into the following constraints on $\eta$

\begin{equation} 
\left[J_i, \eta\right] =0, \quad
\left\{K_i,\eta\right\} =0 .
\end{equation}
Since the only property of the generators of rotations and boosts that enters the derivation of the above constraints is that $\J^\dagger = \J$ and $\K^\dagger = - \K$, the result applies to all \emph{finite} dimensional representations of the Lorentz group. It need not be restricted to Elko alone. Seen in this light, there is no non-trivial solution for $\eta$ either for the $r$-type or the $\ell$-type Weyl spinors.  For $r\oplus\ell$ representation space, the most general solution is found to have the form

\begin{equation}
\eta = 
\left[\begin{array}{cccc}
0 & 0 & a& 0 \\
0 & 0 & 0 & a \\
b & 0 & 0 & 0\\
0 & b & 0 & 0
\end{array}\right] . \label{eq:metric}
\end{equation}
It is now convenient to introduce the notation $e_1(\p) := \xi_{\{-,+\}}(\p)$,  $e_2(\p) := \xi_{\{+,-\}}(\p)$, $e_3(\p) := \zeta_{\{-,+\}}(\p)$, and  $e_4(\p) := \zeta_{\{+,-\}}(\p)$. Sixteen values of
$[e_\imath (\p) ]^\dagger \eta\, e_\jmath(\p)$ as $\imath$ and $\jmath$ vary from 1 to 4 are presented in Table 1.

\begin{table}[htdp]
\caption{The values of $[e_\imath (\p) ]^\dagger \eta\, e_\jmath(\p)$ evaluated using $\eta$.  The $\imath$ runs from 1 to 4 along the rows and $\jmath$ does the same across the columns.}
\begin{center}
\begin{tabular}{|c|c|c|c|}
 \hline
0 & $- i m (a + b)$ & $- i m(a - b)$ & 0 \\
\hline
$i m (a + b)$ & 0 &  0 & $-  i m (a -  b)$\\
\hline
$-  i m (a -  b) $& 0 & 0 &$ i m (a +  b)$\\
\hline
0 & $- i m (a - b) $& $- i m(a + b)$ & 0 \\
\hline
\end{tabular}
\end{center}
\label{dual}
\end{table}
To treat the $r$ and $\ell$ Weyl spaces on the same footing, we set $b=a$.
To make the invariant norms real,  we give $a$ and $b$ the  common value of $\pm i$; resulting in $\eta = \pm i \gamma^0$. Within the stated caveats, the uniqueness of the Elko dual, defined in~Eq.~(\ref{eq:dualchi}), is now apparent.
 
 \subsection{Elko orthonormality and completeness relations}
 
 Under the new dual, the orthonormality  relations read
	\begin{eqnarray}
 		&& {{\stackrel{\neg}{\xi}}_{\alpha}(\p)}\, \xi_{\alpha^\prime}(\p) = 
				+ \,2 m \delta_{\alpha\alpha^\prime},	
		\label{eq:xinorm} \\
        	&& {{\stackrel{\neg}{\zeta}}_{\alpha}(\p)} \,
			\zeta_{\alpha^\prime}(\p) = 
			-\, 2 m \delta_{\alpha\alpha^\prime},
		\label{eq:zetanorm}
	\end{eqnarray}
 along with ${{\stackrel{\neg}{\xi}}_{\alpha}(\p)}\,
 \zeta_{\alpha^\prime}(\p) = 0$, and
 ${{\stackrel{\neg}{\zeta}}_{\alpha}(\p)}\, \xi_{\alpha^\prime}(\p) =
 0 $.  The dual helicity index $\alpha$ ranges over the two
 possibilities: $\{+,-\}$ and $\{-,+\}$, and $-\{\pm,\mp\}:=
 \{\mp,\pm\}$. The completeness relation
	\begin{equation}
	\frac{1}{2 m}\sum_\alpha\big[
 	\xi_{\alpha}(\p)\, {{\stackrel{\neg}{\xi}}_{\alpha}(\p)}-
 	\zeta_{\alpha}(\p)\, {{\stackrel{\neg}{\zeta}}_{\alpha}(\p)}\big] 
	= \openone\label{eq:completeness}
	\end{equation}
 establishes that we need to use {\em both} the self-conjugate as well
 as the anti self-conjugate spinors to fully capture the relevant
 degrees of freedom.
 
 \subsection{Elko spin sums and  a preferred axis}

The existence of a preferred axis, which we will later identify as the axis of locality in the dark sector, is hidden in the spin sums that appear in~Eq.~(\ref{eq:completeness}). It becomes manifest in the results:
	\begin{eqnarray} && \sum_\alpha
	\xi_\alpha(\p) \stackrel{\neg}{\xi}_\alpha(\p) =  m
	\left[ {\mathcal G}(\p) + \openone )\right] ,
	\label{eq:spinsumxi}\\ 
        && \sum_\alpha \zeta_\alpha(\p)
	\stackrel{\neg}{\zeta}_\alpha(\p) = m 
        \left[{\mathcal G}(\p) -\openone\right].\label{eq:spinsumzeta} \end{eqnarray}
 which together {\em define} ${\mathcal G}(\p)$.  A
\textit{direct evaluation} of the
 left hand side of the above equations gives
 \begin{equation} {\mathcal G}(\p)= i \left( \begin{array} {cccc} 0 &
 0 & 0 & - e^{-i \phi}\\ 0 & 0 & e^{i\phi} & 0 \\ 0 & - e^{-i \phi}&0
 & 0\\ e^{i\phi} & 0 & 0 & 0 \end{array} \right). \label{eq:gp}
 \end{equation} 
It is to be  immediately noted that ${\mathcal G}(\p)$ is an odd function of $\p$
 \begin{equation}
 {\mathcal G}(\p) = -\, {\mathcal G}(-\p).\label{eq:gpodd}
 \end{equation} 
 But since ${\mathcal G}(\p)$ is independent of $p$ and
 $\theta$, it is more instructive to translate the above expression into 
 \begin{equation} {\mathcal G}(\phi) = -\, {\mathcal
 G}(\pi+\phi) . \label{eq:Gzimpok}
 \end{equation} 
This serves to define a preferred axis, $z_e$ (see also section \ref{Sec:2.6} below).\footnote{The accompanying $x_e$ and $y_e$ axis help to define a preferred frame.}  Another hint for a preferred axis arises when one notes that the Elko spinorial structure does not  enjoy covariance under usual local $U(1)$ transformation with phase $\exp(i \alpha(x))$. However, $U_E(1) = \exp\left(i\gamma^0\alpha(x)\right)$ ~\textemdash~and not 
$ U_M(1) = \exp\left(i\gamma^5\alpha(x)\right)$ as one would have thought~\cite[p. 72]{Marshak:1969re}~\textemdash~ preserves various aspects of the Elko structure. Similar comments apply to the non-Abelian gauge transformations of the SM.

\subsection{Elko and Dirac spinors: A comparison}
\label{Sec:2.6}
 
For a comparison with the Dirac counterpart, one may define $g^\mu:=(0,\g)$ with
\begin{equation}
\g := - [1/\sin(\theta)] \partial \hp/\partial\phi =
 (\sin\phi,-\cos\phi,0)
  \end{equation}
 Note may be taken that $g^\mu$
 is a
 unit spacelike four-vector, $g_\mu g^\mu = -1$.  Furthermore, $g_\mu p^\mu=0$.
In terms of $g^\mu$, $\mathcal{G}(\p)$ may
 be written as 
 \begin{equation} {\mathcal G}(\p) =
 \gamma^5(\gamma_1\sin\phi - \gamma_2 \cos\phi) = \gamma^5 
 \gamma_\mu g^\mu \label{eq:gp2}
 \end{equation}
This gives Eqs.~(\ref{eq:spinsumxi}) and (\ref{eq:spinsumzeta}),
the form 
	 \begin{eqnarray} \sum_\alpha
	\xi_\alpha(\p) \stackrel{\neg}{\xi}_\alpha(\p)=  m
	\left[ \gamma^5 
           \gamma_\mu g^\mu + \openone \right],
	\label{eq:spinsumxizimpok}\\ 
         \sum_\alpha \zeta_\alpha(\p)
	\stackrel{\neg}{\zeta}_\alpha(\p) = m 
        \left[\gamma^5 
 	\gamma_\mu g^\mu -\openone\right].\label{eq:spinsumzetazimpok} 
	\end{eqnarray}
The appearance of $g^\mu$ on the the right hand side introduces a preferred axis.	
	
The reader is reminded that so far no wave equation has been invoked. The charge conjugation and parity operators can be formally defined without reference to a wave equation. This can be seen from the fact that under parity $\kappa_r \leftrightarrow\kappa_\ell$, and thus the parity operator in the $r\oplus \ell$ representation space equals $\gamma^0$ (modulo a multiplicative  phase factor). Dirac spinors then emerge as eigenspinors of the parity operator. From this perspective, 
when applied to  eigenspinors of the parity operator, charge conjugation interchanges opposite parity eigenspinors (and it takes the form given in Eq.~(\ref{eq:cco})).  Once this view is accepted, one can start with an appropriate counterpart of the Elko at rest and following the same procedure as for Elko obtain the standard Dirac spinors, $u(\p)$ and $v(\p)$. The counterpart of the Elko spin sums then read
	 \begin{eqnarray} \sum_\alpha
	u_\sigma(\p) \overline{u}_\sigma(\p)=  m
	\left[ 
           m^{-1} \gamma_\mu p^\mu + \openone \right],
	\label{eq:spinsumxizimpok-dirac}\\ 
         \sum_\sigma v_\sigma(\p)
	\overline{v}_\alpha(\p) = m 
        \left[ m^{-1}
 	\gamma_\mu p^\mu -\openone\right].\label{eq:spinsumzetazimpok-dirac} 
	\end{eqnarray}
The momentum-space Dirac equations now appear as  identities derived from multiplying
Eq.~(\ref{eq:spinsumxizimpok-dirac}) from the right by $u_{\sigma^\prime}(\p)$,  Eq.~(\ref{eq:spinsumzetazimpok-dirac}) by
 $v_{\sigma^\prime}(\p)$, and using  $\overline{u}_\sigma(\p) {u}_{\sigma^\prime}(\p) = 2 m \delta_{\sigma\sigma^\prime}$ and  $\overline{v}_\sigma(\p) {v}_{\sigma^\prime}(\p) =  - 2 m \delta_{\sigma\sigma^\prime}$.
 That these `identities' are taken to lead to a wave equation, and eventually to derive the  Lagrangian density, may have led to internal inconsistency unless the associated Green function was found to be proportional to $\langle \;\vert \mathcal{T}\left[\Psi(x^\prime) \overline\Psi(x)\right]\vert\;\rangle$, in the usual notation with $\Psi(x)$ as the Dirac quantum field. For the Dirac case this is precisely what happens and no internal inconsistency is introduced by following such a `quick and dirty' route to arrive at the Lagrangian density.
 
 To appreciate these remarks, a similar exercise may be undertaken for Elko. One finds that the resulting identities have no dynamical content.
 
 \subsection{Elko satisfy Klein-Gordon, not Dirac, equation}
 
 The next step in our discourse requires the observation that Elko  do not satisfy the Dirac equation. To see this we apply the operator $\gamma^\mu p_\mu$ on Elko and find the
 following identities 
 	\begin{eqnarray}
  	&&\gp \xi_{\{-,+\}}(\p) = i m \xi_{\{+,-\}}(\p), \label{eq:a} \\
  	&& \gp \xi_{\{+,-\}}(\p) = - i m \xi_{\{-,+\}}(\p) ,\label{eq:b} \\
 	&&\gp \zeta_{\{-,+\}}(\p) = - i m \zeta_{\{+,-\}}(\p) ,
		\label{eq:c} \\
  	&& \gp \zeta_{\{+,-\}}(\p) =  i m \zeta_{\{-,+\}}(\p) .\ \label{eq:d}
 	\end{eqnarray}
  Operating equation (\ref{eq:a})
 from the left by $\gamma^\nu p_\nu$, and then using (\ref{eq:b}) on
 the resulting right hand side, and repeating the same procedure for
 the remaining equations we get
	\begin{eqnarray}
	\left(\gamma^\nu\gamma^\mu p_\nu p_\mu  - m^2\right) 
	\xi_{\{\mp,\pm\}}(\p) = 0 ,\\
	\left(\gamma^\nu\gamma^\mu p_\nu p_\mu  - m^2\right) 
	\zeta_{\{\mp,\pm\}}(\p) = 0 .
	\end{eqnarray}
 Now using $\{\gamma^\mu,\gamma^\nu\} = 2 \eta^{\mu\nu}$, yields the
 Klein-Gordon equation (in momentum space) for the $\xi(\p)$ and
 $\zeta(\p)$ spinors. 
 Aitchison and Hey's  concern ~\cite[Appendix P]{Aitchison:2004cs} is thus overcome. The problem, as is now apparent, resides in the approach
of  constructing  ``simplest candidates for
 a kinematic spinor term''~\cite[p. 34]{Ramond:1981pw}. The latter approach yields the ``correct'' results if Majorana spinors are treated as G-numbers, and the ``wrong'' result if they are treated as c-numbers.  The systematic approach outlined here works in both contexts.

 \subsection{Two quantum fields with Elko as their expansion coefficients}

We now
 examine the physical and mathematical content of two
 quantum fields with $\xi_\alpha(\p)$ and $\zeta_\alpha(\p)$ as their expansion coefficients
	\begin{eqnarray}
 		\Lambda(x) \stackrel{\rm def}{=}&& \int \frac{d^3 p}{(2\pi)^3} 
		\frac{1}{\sqrt{2 m E(\p)}} \sum_\alpha{\Big[}
             	a_\alpha(\p) \xi_\alpha(\p) e^{- i p_\mu x^\mu} \nonumber\\
                   &&\hspace{21pt}+\; b^\ddagger_\alpha(\p) \zeta_\alpha(\p) 
		e^{+ i p_\mu x^\mu}
                 {\Big]}
	\end{eqnarray}
 and
	 \begin{equation}
		\lambda(x)\stackrel{\rm def}{=} 
 \Lambda(x)\big\vert_{b^\ddagger({\bf p}) \to 
						 a^\ddagger({\bf p})} .
	 \end{equation}
 We assume that the annihilation and creation operators satisfy the
 fermionic anticommutation relations
	\begin{eqnarray}
	\{a_\alpha(\p),\;a^\ddagger_{\alpha^\prime}(\p^\prime)\}
	= (2 \pi)^3\, \delta^3(\p-\p^\prime)\,\delta_{\alpha\alpha^\prime},\\
        \{a_\alpha(\p),\;a_{\alpha^\prime}(\p^\prime)\}
	= 0, \quad\{a^\ddagger_\alpha(\p),\;a^\ddagger_{\alpha^\prime}
	(\p^\prime)\}
	=0 .
	\end{eqnarray}
 Similar anticommutators are assumed for the $b_\alpha(\p)$ and
 $b^\ddagger_\alpha(\p)$. The adjoint field
 $\stackrel{\neg}{\Lambda}(x)$ is defined as
	\begin{eqnarray}
 		\stackrel{\neg}\Lambda(x) \stackrel{\rm def}{=} &&
                \int \frac{d^3 p}{(2\pi)^3} 
		\frac{1}{\sqrt{2 m E(\p)}} \sum_\alpha{\Big[}
             	a^\ddagger_\alpha(\p) \stackrel{\neg}{\xi}_\alpha(\p) 
	        e^{+ i p_\mu x^\mu} \nonumber\\
                   &&\hspace{21pt}+\; b_\alpha(\p) \stackrel{\neg}{\zeta}_\alpha(\p) 
		e^{- i p_\mu x^\mu}
                 {\Big]}
	\end{eqnarray}
The results contained in Eqs.~(\ref{eq:a}-\ref{eq:d}) assure us that it is the Klein-Gordon, and not the Dirac, operator that annihilates the fields $\Lambda(x)$ and $\lambda(x)$.  The associated Lagrangian densities are 
	\begin{eqnarray}
                   &&{\mathcal L}^\Lambda(x) = 
         \partial^\mu\stackrel{\neg}{\Lambda}(x)
                   \partial_\mu\Lambda(x) - m^2
		   \stackrel{\neg}{\Lambda}(x)\Lambda(x),\label{eq:Lagrangian}
		   \\
		  && 
                   {\mathcal L}^\lambda(x) = {\mathcal L}^\Lambda(x) \big\vert_{\Lambda\rightarrow\lambda}
	\end{eqnarray}
The mass dimensionality  of these Elko fields  is thus one, and not three half.  Green functions and the consistency of these result with 
$
\langle \;\vert \mathcal{T}[\Lambda(x^\prime) \stackrel{\neg}{\Lambda}(x)]\vert\;\rangle \;\mbox{and} \;\langle\; \vert \mathcal{T}[\lambda(x^\prime) \stackrel{\neg}{\lambda}(x)]\vert\;\rangle
$
 shall be reported in an archival publication.

To study the locality structure of the fields $\Lambda(x)$ and $\lambda(x)$, we observe that field momenta  are
	\begin{equation}
		\Pi(x)= \frac{\partial{\mathcal L}^\Lambda}
		{\partial\dot\Lambda} =
		\frac{\partial}{\partial t}\stackrel{\neg}{\Lambda}(x),\quad
\end{equation}
 and similarly $ \pi(x) = \frac{\partial}{\partial
 t}\stackrel{\neg}{\lambda}(x)$.  The calculational details for the
 two fields now differ significantly. We begin with the evaluation of
 the equal time anticommutator for  $\Lambda(x)$ and its conjugate
 momentum
	\begin{eqnarray} &&\{\Lambda(\x,t),\; \Pi(\x^\prime,t)\} =
	i\int\frac{d^3 p}{(2\pi)^3}\frac{1}{2 m} e^{i {\mathbf p}\cdot
	({\mathbf x}-{\mathbf x}^\prime)} \nonumber\\
	&&\times\underbrace{\sum_\alpha\left[ \xi_\alpha(\p)
	\stackrel{\neg}{\xi}_\alpha(\p) - \zeta_\alpha(- \p)
	\stackrel{\neg}{\zeta}_\alpha(- \p)\right]}_{=\,2 m [\openone +
	{\mathcal G}(\mathbf{p})]} .\nonumber 
	\end{eqnarray}
The term containing
 ${\mathcal G}(\p)$ vanishes only when ${\mathbf x}-{\mathbf x}^\prime$ lies along the $z_e$ axis (see Eq.~(\ref{eq:gpodd}), and discussion of this integral in Ref.~\cite{Ahluwalia:2004sz,Ahluwalia:2004ab})
	\begin{equation}
		{\mathbf x}-{\mathbf x}^\prime \;\mbox{along }z_e:\quad \{\Lambda(\x,t),\; \Pi(\x^\prime,t)\} 
		=  i \delta^3(\x -\x^\prime) \openone.\label{eq:LPac}
	\end{equation}
The anticommutators for the particle/antiparticle annihilation and creation
operators suffice to yield the remaining locality conditions,
	\begin{equation}
		\{\Lambda(\x,t),\; \Lambda(\x^\prime,t)\} =
		\mathbb{O},\quad \{\Pi(\x,t),\; \Pi(\x^\prime,t)\}
		\label{eq:LLPPac} = \mathbb{O}.
	\end{equation}
The set of anticommutators contained in Eqs.~(\ref{eq:LPac}) and
(\ref{eq:LLPPac}) establish that $\Lambda(x)$ becomes local along the $z_e$ axis. For this reason we call $z_e$ as the dark axis of locality.

 For the equal time anticommutator of the $\lambda(x)$ field with its
 conjugate momentum, we find
	\begin{eqnarray}
		&&\{\lambda(\x,t),\; \pi(\x^\prime,t)\} =  
		i\int\frac{d^3 p}{(2\pi)^3}\frac{1}{2 m}
		\nonumber\\
		&&\times\sum_\alpha\left[ e^{i {\mathbf p}\cdot
		({\mathbf x}-{\mathbf x}^\prime)}
		\left(\xi_\alpha(\p) \stackrel{\neg}{\xi}_\alpha(\p) -
		\zeta_\alpha(- \p) 
		\stackrel{\neg}{\zeta}_\alpha(- \p)\right)\right].\nonumber
	\end{eqnarray}
 Which, using similar arguments as before, yields
	\begin{equation}
	{\mathbf x}-{\mathbf x}^\prime \;\mbox{along } z_e:\quad
		\{\lambda(\x,t),\; \pi(x^\prime,t)\} 
		=  i \delta^3(\x -\x^\prime) \openone  .\label{eq:lpac2}
	\end{equation}
 The difference arises in the evaluation of the remaining
 anticommutators.  The equal time $\lambda$-$\lambda$ anticommutator
 reduces to
	\begin{eqnarray} 
		&&\{\lambda(\x,t),\; \lambda(\x^\prime,t)\} =
	 	\int\frac{d^3 p}{(2\pi)^3}\frac{1}{2 m E(\p)}\; e^{i {\mathbf
		p}\cdot ({\mathbf x}-{\mathbf x}^\prime)} \nonumber\\
		&&\times\underbrace{\sum_\alpha\left[ \xi_\alpha(\p)
		\zeta^T_\alpha(\p) + \zeta_\alpha(- \p) \xi^T_\alpha(-
		\p)\right]}_{=:\,\Omega(\p)}.  \label{eq:zimpokb}
	\end{eqnarray}
 Now using explicit expressions for $ \xi_\alpha(\p)$ and $
 \zeta_\alpha(\p)$ we find that $\Omega(\p)$ identically vanishes.
 Equation~(\ref{eq:zimpokb}) then implies
 	\begin{equation}
		\{\lambda(\x,t),\; \lambda(\x^\prime,t)\} =  \mathbb{O}.          
		\label{eq:llac2}
	\end{equation}
 And, finally the equal time $\pi$-$\pi$ anticommutator simplifies to
	\begin{eqnarray}
		&&\{\pi(\x,t),\; \pi(\x^\prime,t)\} = \int\frac{d^3
		p}{(2\pi)^3}\frac{E(\p)}{2 m}\; e^{-i {\mathbf p}\cdot
		({\mathbf x}-{\mathbf x}^\prime)} \nonumber\\
		&&\times\underbrace{\sum_\alpha\left[
		\Big(\stackrel{\neg}{\xi}_\alpha(\p) \Big)^T
		\stackrel{\neg}{\zeta}_\alpha(\p) +
		\Big(\stackrel{\neg}{\zeta}_\alpha(-\p) \Big)^T
		\stackrel{\neg}{\xi}_\alpha(-\p)
		\right]}_{=\mathbb{O},~ \mbox{by a direct evaluation
		}},\nonumber
	\end{eqnarray}
 yielding
	\begin{equation}
		\{\pi(\x,t),\; \pi(\x^\prime,t)\} =
		\mathbb{O}. \label{eq:ppac2}
	\end{equation}
Again, $\lambda(x)$ becomes local along $z_e$. This further justifies the term  `dark axis of locality' for the $z_e$ axis.

 The dimension four interactions of the $\Lambda(x)$ and $\lambda(x)$
 with the standard model fields are restricted to those with the SM
 Higgs doublet $\phi(x)$. These are
	\begin{equation}
        	  {\mathcal L}^{\mathrm{int}}(x) =
        	  \phi^\dagger(x)\phi(x)\,\sum_{\psi,\Psi}
        	  a_{\psi\Psi}\stackrel{\neg}{\psi}(x) {\Psi}(x),
	\end{equation}
 where $a_{\psi\Psi}$ are unknown coupling constants and
 symbols $\psi$ and $\Psi$ stand for either $\Lambda$ or $\lambda$.
 By virtue of their mass dimensionality the new Elko fields are
 endowed with dimension four quartic self interactions contained in
	\begin{equation} 
	 	{\mathcal L}^{\mathrm{self}}= \sum_{\psi,\Psi}
	 	b_{\psi\Psi} \left[\stackrel{\neg}{\psi}(x)
	 	{\Psi}(x)\right]^2,
	\end{equation}
 where $b_{\psi\Psi}$ are unknown coupling constants.
 
 Remarks following Eq.~(\ref{eq:Gzimpok}) suggest that the
 Elko fields need not be self referentially dark.
 However, the same remarks imply that  quantum fields based on Elko may not participate
 in interactions with the standard model gauge fields. This also  allows the Elko-based  dark matter to evade the constraints on preferred-frame effects discussed in literature (see, e.g., 
 Ref.~\cite{Collins:2004bp}).
 
 \section{Concluding remarks}  
 
 This paper is a natural and nontrivial continuation of the 
 2005 work of Ahluwalia and  Grumiller on Elko.  Here we reported that Elko breaks Lorentz symmetry in a rather subtle and unexpected way by containing a `hidden' preferred direction.  Along this preferred direction, a quantum field based on Elko enjoys locality. 
In the form reported here, Elko offers  mass dimension one fermionic dark matter with a quartic self-interaction and a preferred axis of locality. The locality result crucially depends on a judicious choice of phases.

 \section*{Acknowledgments}

  We thank Adam Gillard and Ben Martin for discussions, and also Karl-Henning Rehren
 for his helpful comments. 

\end{document}